%% file: main.tex
\documentclass[
]{ceurart}

\usepackage{listings}
\usepackage{cleveref}
\usepackage{subcaption}
\begin{document}

\copyrightyear{2026}
\copyrightclause{Copyright © 2026 for this paper by its authors. Use permitted under Creative Commons License Attribution 4.0 International (CC BY 4.0).}
\conference{FRAME'26: Methodology First - Rethinking Research Assessment in RecSys Workshop, September 28, 2026, Minneapolis, Minnesota, USA}
\title{SPADE: Escaping the Popularity-Similarity Frontier to Measure Serendipitous Recommendations}

%
\author[1]{Tobias Vente}[%
orcid=0009-0003-8881-2379,
email=tobias.vente@uantwerpen.be,
url=https://tobiasvente.github.io/,
]

\author[1]{Maarten Peirsman}[%
orcid=0009-0004-1394-354X,
email=maarten.peirsman@uantwerpen.be,
]

\author[1]{Noah Dani\"els}[%
orcid=0009-0001-7778-4769,
email=noah.daniels@uantwerpen.be,
]

\author[2]{Hannu Toivonen}[%
orcid=0000-0003-1339-8022,
email=hannu.toivonen@helsinki.fi,
url=http://www.cs.helsinki.fi/hannu.toivonen/,
]

\author[1,3]{Bart Goethals}[%
orcid=0000-0001-9327-9554,
email=bart.goethals@uantwerpen.be,
url=https://www.bartgoethals.com/,
]

\address[1]{ADREM Data Lab, University of Antwerp,
  Middelheimlaan 1, 2020 Antwerpen, Belgium}
\address[2]{Department of Computer Science, PO Box 68, FI-00014 University of Helsinki, Finland}
\address[3]{Froomle, Post X, Borsbeeksebrug 34, 2600 Antwerpen, Belgium}

\input{content/abstract}

\maketitle

\input{content/introduction}
\input{content/related_work}
\input{content/spade}
\input{content/experimental_setup}
\input{content/results}

\bibliography{sample-ceur}

\end{document}

%% file: content/abstract.tex
\begin{abstract}
Recommender systems engineer serendipity to foster active exploration and break predictable consumption cycles. 
The problem with existing offline beyond-accuracy metrics is that they often either isolate historical similarity or global popularity. 
We aim to design an evaluation metric that examines similarity, popularity, and actual user relevance. 
To achieve this, we introduce SPADE (Serendipitous Pareto Distance Evaluation). SPADE maps all items into a two-dimensional space to directly calculate a user-specific Pareto frontier of maximally popular and historically similar items. 
The final serendipity score is then computed by averaging the minimum Euclidean distance from this boundary strictly for the correctly recommended test-set items. 
Evaluating SPADE across five datasets and five baseline algorithms confirms its effectiveness; our results show that the metric successfully prevents algorithms from exploiting beyond-accuracy measures with irrelevant or non-personalized recommendations, reliably isolating serendipitous discoveries.
\end{abstract}

%% file: content/introduction.tex
\section{Introduction}
Recommender systems excel at predicting user preferences, yet they often struggle to deliver what users never knew they needed \cite{10.1145/1864708.1864761, ziarani2021serendipity}. 
When systems strictly optimize for predictive accuracy, they risk trapping users in isolated information silos and continuously narrowing their exposure to diverse content \cite{10.1145/2566486.2568012}. 
To overcome this filter bubble effect and ensure long-term user retention, researchers engineer serendipity into these systems \cite{10.1145/3711896.3737199}. 
Serendipity acts as a counterweight, fostering active exploration and breaking predictable consumption cycles \cite{10.1145/1864708.1864761}.

While dictionary definitions of serendipity emphasize pure chance and luck, in the context of recommender systems, it is treated instead as an explicit engineering objective.
Serendipity is defined as the intersection of surprise, relevance, and novelty in recommendations \cite{10.1145/2792838.2800184, 10.1145/2926720, 10.1145/3167132.3167276}. 
Introducing serendipity does not mean serving random items; rather, 
it involves delivering suggestions that appear as fortunate, random discoveries to the user, while in fact being purposefully selected by the algorithm \cite{10.1145/2792838.2800184, 10.1145/2926720, 10.1145/3167132.3167276}. 
For example, recommending the next obvious blockbuster or a direct sequel almost guarantees high acceptance, but it fails to deliver a surprising discovery \cite{ziarani2021serendipity}. 
In contrast, recommending an item from an unexpected yet relevant genre achieves serendipity \cite{jenders2015serendipity}.

The degree of serendipity in a recommendation relies on two distinct dimensions: its similarity to the target user's historical interactions and its global popularity across the system \cite{10.1145/2926720, 10.1145/2124295.2124300}. 
While suggesting highly similar or globally dominant items is a safe strategy to ensure baseline relevance, it fundamentally compromises both unexpectedness and novelty \cite{10.1145/2926720, 10.1145/2124295.2124300}. 
A recommendation that too closely mirrors a user's past interactions fails to generate surprise, just as a universally popular item is highly unlikely to provide a novel discovery. 
Consequently, achieving serendipity requires navigating away from these high-probability zones to find items that are relevant, yet neither popular nor similar to a user's past interactions.

Unfortunately, established serendipity metrics struggle to capture this dynamic due to their narrow focus. They tend to evaluate serendipity by isolating just one dimension of the recommendation space. For example, some approaches measure unexpectedness purely as the content-based distance between recommendations and a user's history \cite{shani2010evaluating, de2014comparison, wang2018serendipitous}. Other metrics focus entirely on global patterns. These include evaluating serendipity as a deviation from primitive popularity baselines \cite{murakami2007metrics, 10.1145/1864708.1864761}, relying on item co-occurrence probabilities \cite{kaminskas2014measuring, kaminskas2016diversity}, or measuring the time saved in discovering an item \cite{kawamae2015real}.

While existing serendipity metrics capture distinct facets of serendipity, they share a flaw: they do not assess both similarity and popularity simultaneously. 
By ignoring either similarity or popularity, isolated metrics often reward mainstream items simply because they are new to the user, or reward highly predictable items just because they are globally rare, or inadvertently reward obscure discoveries that hold no actual relevance to the user.
Failing to penalize these false positives can inflate the measured serendipity, exposing the need for a metric that strictly optimizes for both dimensions.

To address this gap, we introduce SPADE (Serendipitous Pareto Distance Evaluation), a novel offline beyond-accuracy metric that quantifies serendipity by jointly evaluating historical similarity and global popularity. 
Within this two-dimensional space, we construct a user-specific Pareto frontier of items that maximize both global reach and alignment with a user's history. SPADE measures serendipity as the Euclidean distance from an algorithm's correctly predicted items to the closest point on the Pareto front. 
By defining SPADE and evaluating it against established beyond accuracy metrics, we investigate whether this structural approach can successfully measure serendipity. Ultimately, SPADE provides researchers with a standard, quantifiable method to determine whether an algorithm is merely retreating to predictable and popular content or if it is actively delivering serendipitous discoveries.

To validate SPADE, we execute three research tasks: formalization, offline evaluation, and comparative analysis. 
First, we define SPADE's mathematical framework, detailing the construction of the Pareto frontier and the calculation of distances. 
Then, by comparing SPADE against commonly used beyond-accuracy metrics (Novelty, Co-Occurrence, and Primitivity) across five baseline algorithms on five widely used datasets, we demonstrate how it isolates serendipitous items that traditional metrics misclassify.

The code used to execute the experiments is publicly available in our GitHub\footnote{\url{https://github.com/Maarten11/SPADE}} repository and further contains documentation to ensure the reproducibility of our experiments

%% file: content/related_work.tex
\section{Related Work}
\label{sec:rel_work}
While the importance of serendipity in recommender systems is widely recognized, existing evaluation metrics largely approach the concept through isolated dimensions. A review of the literature reveals three primary methodologies for quantifying serendipity and unexpectedness offline. However, because these approaches rely on a narrow focus (prioritizing either user history, primitive baselines, or temporal distance), they struggle to capture the necessary dual dynamic of historical similarity and global popularity, leaving them vulnerable to false positives.

The most frequent approach to measuring serendipity calculates the distance between a recommended item and the user’s historical profile, utilizing either content features or collaborative filtering (CF) metrics. Initially outlined by Shani and Gunawardana \cite{shani2010evaluating}, content-based distance measurements have been widely adapted to evaluate how far a new recommendation deviates from a user's known preferences \cite{de2014comparison, jelassi2015towards, wang2018serendipitous}. Variations of this metric, which calculate the difference between recommendations and user profiles through slightly different formulas, have been used to evaluate user exploration \cite{chen2021values} and multi-objective diversity \cite{sa2022diversity}. 

Similarly, other researchers measure unexpectedness by calculating CF-based distances, frequently employing Point-wise Mutual Information (PMI) or Normalized PMI (NPMI) to evaluate the statistical co-occurrence between a candidate item and the specific items in a user's historical profile \cite{kaminskas2014measuring, kaminskas2016diversity, kaya2022novel}. While effective at measuring personalization, these profile-centric metrics are entirely blind to global item popularity. By focusing solely on the user's history, they risk rewarding highly popular items simply because they are absent from a specific user's profile, falsely categorizing a highly predictable global trend as a serendipitous discovery.

A second prominent approach operationalizes serendipity by comparing a proposed algorithm's outputs against those of a ``primitive'' recommender system. Originally introduced by Murakami et al. \cite{murakami2007metrics} and famously formalized by Ge et al. \cite{10.1145/1864708.1864761}, this metric conceptualizes serendipity as the proportion of relevant items that are unexpected. Unexpectedness is defined by the items the primitive baseline fails to recommend. This method has been widely adopted across various domains, including news recommendations \cite{maksai2015predicting}, serendipity-oriented greedy algorithms \cite{kotkov2017serendipity}, fuzzy inference systems \cite{babu2015evaluation}, research paper recommendations \cite{nishioka2019towards, nishioka2020influence}, offline data processing \cite{ren2015survey}, and deep neural networks \cite{adamopoulos2014unexpectedness, ziarani2021deep}. The fundamental flaw of this approach is its heavy reliance on the chosen baseline. Because the primitive system can be based on either global popularity or historical similarity (recommending the same content), the resulting serendipity score is ambiguous. It cannot guarantee that an item is both a long-tail discovery and a taste-broadening deviation.

A more specialized metric defines serendipity through the lens of time. Introduced by Kawamae \cite{kawamae2010serendipitous}, this approach measures the difference between the time an algorithm recommends an item and the time it would theoretically take a user to find that item on their own without the system's help. An item with a large time difference is considered highly serendipitous because it implies the user would not have naturally discovered it in the near future. This metric has been primarily used to evaluate real-time recommendations \cite{kawamae2015real, wang2018serendipitous}. While conceptually interesting, this temporal approach requires extensive, time-stamped historical logs of organic user discovery, which are rarely available. Furthermore, it measures the symptom of serendipity (delayed discovery) rather than the underlying structural causes—the item's relationship to the user's taste and the global market.

While our work introduces the Pareto front as an evaluation tool, the concept itself is well-established in recommender systems literature, primarily within the domain of multi-objective optimization \cite{lin2019pareto, ribeiro2014multiobjective}. Because recommender systems often face competing goals—such as maximizing accuracy while maintaining catalog coverage, diversity, or fairness—researchers frequently employ Pareto frontiers to find optimal trade-offs during model training \cite{li2024deep, ribeiro2014multiobjective}. For instance, algorithms are often designed to push recommendations toward a Pareto-optimal boundary where accuracy cannot be further improved without sacrificing diversity \cite{xiao2017fairness, peng2024reconciling}.

Critically, the existing literature uses Pareto frontiers almost exclusively as an optimization objective for generating recommendations rather than as a post-hoc evaluation metric. 

Ultimately, the current literature lacks a unified metric that evaluates serendipity as a function of both user-level similarity and global popularity. Because existing metrics ignore this crucial trade-off, they lack a reliable mechanism to penalize the false positives that occur at the extremes of either dimension.

%% file: content/spade.tex
\section{SPADE (Serendipitous Pareto Distance Evaluation)}

To evaluate serendipity, every candidate item $i$ for a user profile $U$ is mapped into a two-dimensional space defined by its global popularity and its similarity to the user's profile. We calculate a user-specific Pareto front of items that maximize both of these dimensions. Each item on this front is either the most popular item among items that are at least as similar or the most similar item among items that are at least as popular. We posit that these Pareto-optimal items are the least serendipitous because they dominate either popularity or similarity. Conversely, items farther from the Pareto front are considered more serendipitous, as they are simultaneously less popular and less similar to items on the front 
(\cref{fig:user_1118}).
To compute the final SPADE score, we calculate the average distance from the Pareto front only for the correctly recommended items (i.e., those in the user's test set).

\subsection{Defining the Dimensions}

For this Pareto formulation to be well-posed, the two dimensions must represent a genuine trade-off. It is therefore critical to choose a similarity measure that is not positively correlated with popularity.
If a positively correlated measure such as cosine similarity is used~\cite{lin2024recommendation}, the most popular items will inherently also score as the most similar. Geometrically, this causes the candidate items to cluster along a single diagonal, and the Pareto front collapses into a single point (or a very narrow band) in the extreme upper-right of the space. When the front collapses in this way, there is no longer a boundary of trade-offs, and measuring distance to the front becomes meaningless. We therefore require opposing measures to scatter items across the space, forcing a wide, distinct curve that represents the maximum possible trade-off between popularity and similarity.

\textbf{Popularity ($x$-axis):} We define the popularity of a candidate item $i$ by its interaction frequency, normalized to a range of $[0, 1]$. During training, the occurrence $n_i$ of each item is counted and divided by the maximum item count across the entire item set $I$:
$$pop(i) = \frac{n_i}{\max_{k \in I} n_k}$$

\textbf{Similarity ($y$-axis):} To satisfy the independent requirement, we utilize \emph{unnormalized} pointwise mutual information (PMI). The PMI of two items is calculated based on the fraction of users who interacted with either and both items as follows:
$$\text{PMI}(i,j)=\log \dfrac{\frac{n_\mathit{ij}}{n}}{\frac{n_\mathit{i}}{n}\frac{n_\mathit{j}}{n}} = \log \dfrac{n_\mathit{ij}}{\frac{1}{n}n_\mathit{i}n_\mathit{j}}$$
where $n_\mathit{i}$ and $n_\mathit{j}$ indicate the number of users who have interacted with items $i$ and $j$ respectively, $n_\mathit{ij}$ represents the number of users who interacted with both items, and $n$ indicates the total number of users. It is well known that PMI is negatively correlated with popularity~\cite{Bouma2009NormalizedM}, making it ideal for our use case. We intentionally avoid Normalized PMI~\cite{kaminskas2014measuring, kaminskas2016diversity}, as the normalization process reintroduces positive correlation with popularity.

To counteract PMI's sensitivity to noise for rare items, we apply empirical Bayes smoothing~\cite{robbins1992empirical}. This dampens the effect of low counts and shifts PMI towards 0 when evidence for the item pair is small. We also clip the similarities to be strictly non-negative, as negative similarities predominantly represent noise. The formula for our smoothed Positive PMI (PPMI) is:
$$
\text{PPMI}_\alpha(i,j)=\max\left\{0, \log\dfrac{n_\mathit{ij}+\alpha}{\frac{1}{n}n_\mathit{i} n_\mathit{j}+\alpha}\right\}
$$
We fix the smoothing strength at $\alpha=1$ in our experiments. To compute the overall similarity between a candidate item $i$ and the user profile $U$, we aggregate the pairwise similarities by taking the sum over all items $j$ in the user's history:
$$sim(i, U) = \sum_{j \in U} \text{PPMI}_\alpha(i,j)$$

\subsection{Space Normalization and Distance Calculation}

Because popularity and profile similarity naturally operate on different scales, calculating a distance metric would result in the larger-scaled axis dominating the space. To ensure both dimensions contribute equally, we apply Min-Max scaling to the profile similarities for each user, projecting the similarity axis into the same $[0, 1]$ range as the popularity axis. We denote this scaled similarity as $\widetilde{sim}(i, U)$.

With both axes bounded in $[0, 1]$, we formally define the user-specific Pareto front $\mathcal{P}_U$. An item $p$ belongs to $\mathcal{P}_U$ if there exists no other item $k$ such that both $pop(k) > pop(p)$ and $\widetilde{sim}(k, U) > \widetilde{sim}(p, U)$ hold true. 

Finally, the SPADE score of a candidate item $i$ is calculated as the minimum Euclidean distance $i$ to the Pareto front $\mathcal{P}_U$:
$$\text{SPADE}(i, U) = \min_{p \in \mathcal{P}_U} \sqrt{(pop(i) - pop(p))^2 + (\widetilde{sim}(i, U) - \widetilde{sim}(p, U))^2}$$
A higher SPADE score indicates that an item is located further away from the Pareto-optimal boundaries of popularity and similarity, effectively quantifying its serendipitous value.

The score for a user's recommendation $R_U$ is calculated as the average SPADE score over all correctly recommended items (i.e., the items in user $U$'s test set). 
To calculate the SPADE score for an algorithm, we average the scores over all the users in the test set.


%% file: content/experimental_setup.tex
\section{Experimental Setup}
Our experiments demonstrate the effectiveness of SPADE in evaluating recommendation serendipity relative to popular serendipity metrics. 
We evaluate SPADE using baseline recommendation algorithms and common dataset configurations reported in the recommender systems literature, ensuring that SPADE's performance can be applied to general recommender system research. 
Our analysis focuses on the top-N recommendation task, specifically examining how accurately the metric isolates genuinely unexpected items from easily predictable false positives based entirely on offline data.

\textbf{Datasets:}
We evaluate SPADE across five datasets from distinct domains: CiteULike \cite{DBLP:conf/ijcai/WangCL13}, 
MIND \cite{wu2020mind}, MovieLens-1M \cite{10.1145/2827872}, MovieLens-20M \cite{10.1145/2827872}, 
and Netflix\footnote{\url{https://www.kaggle.com/datasets/netflix-inc/netflix-prize-data}}. 
\Cref{table:dataset-stats} shows statistical information on the datasets. 

\begin{table}[h]
\centering
\begin{tabular}{lrrrrrr}
\toprule
Dataset & \# Interactions & \# Users & \# Items & Avg.\#Int. per user & Avg.\#Int. per item & Sparsity \\
\midrule
CiteULike & 200,251 & 5,550 & 15,439 & 36.08 & 12.97 & 99.77\% \\
MIND* & 1,086,939 & 10,000 & 33,220 & 108.69 & 32.72 & 99.67\% \\
ML-1M & 574,385 & 6,038 & 3,125 & 95.13 & 183.80 & 96.96\% \\
ML-20M* & 730,034 & 10,000 & 11,361 & 73.00 & 64.26 & 99.36\% \\
Netflix* & 1,234,880 & 10,000 & 15,093 & 123.49 & 81.82 & 99.18\% \\
\bottomrule
\end{tabular}
\caption{Statistical information on the datasets used in our experiments. Datasets marked with (*) have been subsampled to 10,000 users to keep within hardware limitations.}
\label{table:dataset-stats}
\end{table}

\textbf{Algorithms:} We compute SPADE against recommendations predicted by five baseline algorithms: EASE, SLIM, Item-based k-Nearest Neighbors (ItemKNN), Popularity, and Random. 
We use the algorithm implementations from the Recpack library \cite{michiels2022recpack} and perform a grid search to tune the hyperparameters\footnote{More details, like the search spaces, can be found in our GitHub repository: \url{https://github.com/Maarten11/spade}}. 


\textbf{Preprocessing}
\label{section:preprocessing}
Following common practice, we perform 5-core filtering across all datasets, pruning the data to ensure that each item contains at least 5 interactions \cite{barkan2021anchor, melchiorre2022protomf}. 
For the most intensive data sets, MIND, MovieLens-20M and Netflix, we subsample the user populations due to hardware memory limitations, capping them at 10,000 users.

\textbf{Beyond-Accuracy Metrics:}
To evaluate the effectiveness of SPADE against existing standards, our experimental setup uses three established baseline metrics that align directly with the methodologies discussed in \cref{sec:rel_work}.
First, to represent the profile-centric evaluation approaches, we measure \textbf{Co-Occurrence}. 
This metric operationalizes collaborative filtering (CF) distance between a candidate item and a user's historical profile, using Point-wise Mutual Information (PMI) to assess statistical unexpectedness.
Second, to capture the comparative baseline methodology, we evaluate \textbf{Primitivity}. 
Following the framework established by Ge et al., this metric calculates serendipity as the proportion of recommendations that deviate from the outputs of a ``primitive'' system (which, for our experiments, is defined as a standard popularity-based recommender).
Finally, we include \textbf{Novelty}, calculated as the inverse global popularity of the recommended items. Because profile-distance metrics (like Co-Occurrence) are often blind to global trends, Novelty serves as the traditional proxy for uncovering long-tail items, allowing us to independently isolate the global popularity dimension that SPADE structurally integrates.

\textbf{Pipeline:}
For our evaluation, we rely on the strong generalization data-split scenario\footnote{\url{https://recpack.froomle.ai/generated/recpack.scenarios.StrongGeneralization.html\#recpack.scenarios.StrongGeneralization}} utilizing an 80/20 train-test split, whereas 20\% of the train set is used as validation data. We evaluate both standard recommendation performance and all included metrics at cut-offs $K \in \{1, 2, 3, 5, 10, 20, 30\}$.

\textbf{Hardware:}
All experiments and evaluations were executed on an Apple Silicon M5 Pro with 64 GB of RAM.




%% file: content/results.tex
\section{Results \& Discussion}
In this section, we empirically evaluate the efficacy of the SPADE metric at both the user and dataset levels. We first provide a qualitative validation of its underlying mechanics through specific user case studies, followed by a comprehensive cross-domain quantitative analysis comparing SPADE against popular beyond-accuracy metrics.

\subsection{Qualitative Validation: The Mechanics of SPADE}
To illustrate the functionality and practical value of the SPADE metric, we examine the recommendation space for two distinct users who show serendipitous item interactions (User 1874 and User 2186). 
For both users, we visualize the item space by mapping normalized similarity to the user's historical profile on the $y$-axis and global item popularity on the $x$-axis. 
In these visualizations, candidate items are represented by blue dots, predictions generated by the EASE baseline are marked in green, and ground truth interactions from the test set are denoted by red crosses. The Pareto front—representing the optimal trade-off between similarity and popularity—is delineated by a solid line.

\begin{figure}[htbp]
    \centering
    
    \begin{subfigure}[b]{0.49\textwidth}
        \includegraphics[width=\textwidth]{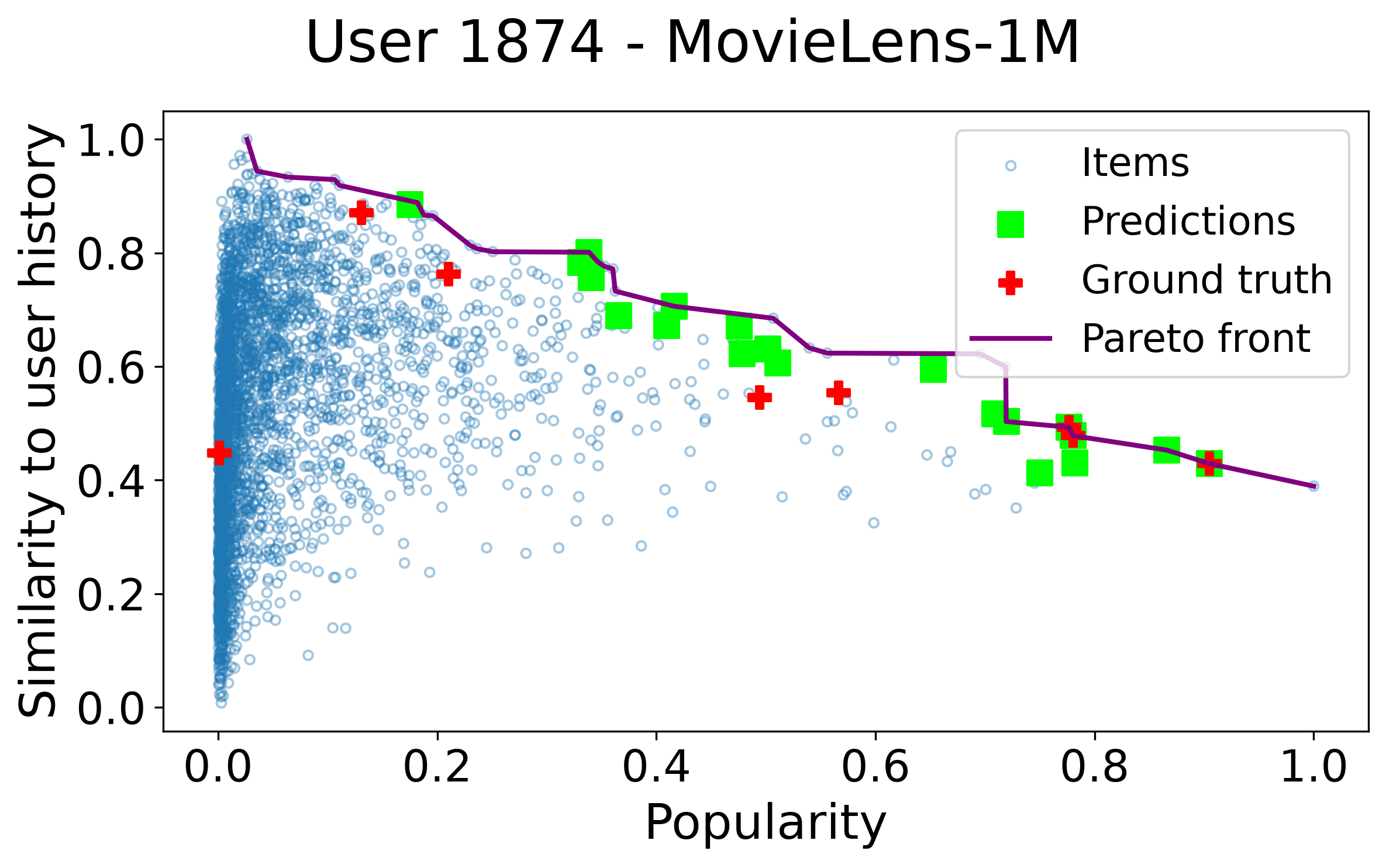}
        \label{fig:user_1118}
    \end{subfigure}
    \hfill
    \begin{subfigure}[b]{0.49\textwidth}
        \includegraphics[width=\textwidth]{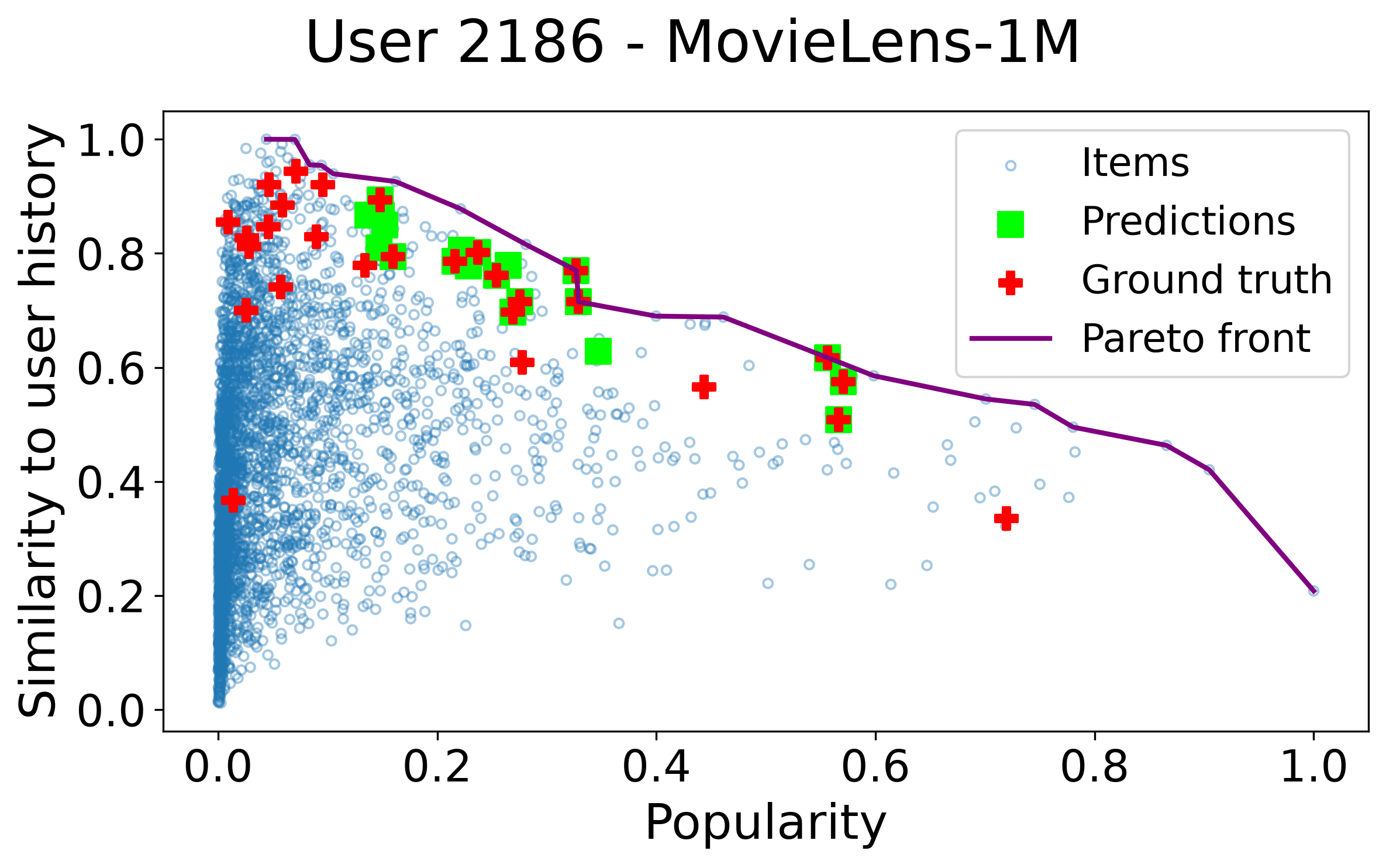}
        \label{fig:user_272}
    \end{subfigure}
    
    \label{fig:my_stacked_images}
    \caption{The $y$-axis denotes normalized item similarity to the user's historical profile, and the $x$-axis denotes global item popularity. Candidate items are represented by blue dots, EASE baseline predictions in green, and ground truth interactions (test set) by red crosses. The solid line indicates the Pareto front maximizing similarity and popularity. More serendipitous items are in the bottom-left quadrant. These relevant but unexpected items lie furthest from the Pareto front and therefore receive the highest serendipity scores under the SPADE metric.}
\end{figure}

In both cases, we observe a ground truth interaction that deviates significantly from the user's typical behavior, falling far outside the high-similarity, high-popularity cluster near the Pareto front.

\textbf{Case Study 1 - User 1874:} For User 1874, one test-set item stands out in the bottom-left quadrant of the plot in \cref{fig:user_1118}: the 1997 Sci-Fi movie Retroactive. 
This user's historical interactions are dominated by Drama, Crime, and Thriller films, with Sci-Fi movies accounting for only 9\% of their viewing history. 
Correspondingly, the item exhibits a low similarity score of 0.46 and an exceptionally low popularity score of 0.005. 
Despite being neither popular globally nor aligned with the user's established preferences, the user still interacted with it, proving its underlying relevance. 
Because this item lies the furthest distance from the Pareto front, SPADE identifies it as serendipitous and assigns it the highest serendipity score. This demonstrates SPADE's ability to reward the successful recommendation of deeply serendipitous, relevant items.

\textbf{Case Study 2 - User 2186:} User 2186 provides a contrasting example of serendipity that is even more extreme but equally valid. 
The standout ground truth item here is the 2000 crime movie The Way of the Gun. 
While this user primarily consumes Action and Adventure, Crime movies constitute 11.6 \% of their interaction history. 
This item yields a similarity score of 0.38 and a popularity score of 0.05. 
If a recommender system were to successfully predict it, SPADE would award it the highest serendipity score for this user, acknowledging that the recommendation pushes the boundaries of their typical consumption habits while remaining relevant.

Ultimately, these examples validate the core premise of SPADE: by measuring distance from the Pareto front, the metric successfully identifies and rewards items that break away from obvious popularity and historical-similarity biases, capturing serendipity.

\subsection{Cross-Domain Empirical Evaluation}

Having established the foundational mechanics of the SPADE metric at the user level, we extend our evaluation to a cross-domain quantitative analysis to assess its robustness compared to traditional beyond-accuracy metrics.

\begin{figure}[htbp]
    \centering
    \includegraphics[width=1\textwidth]{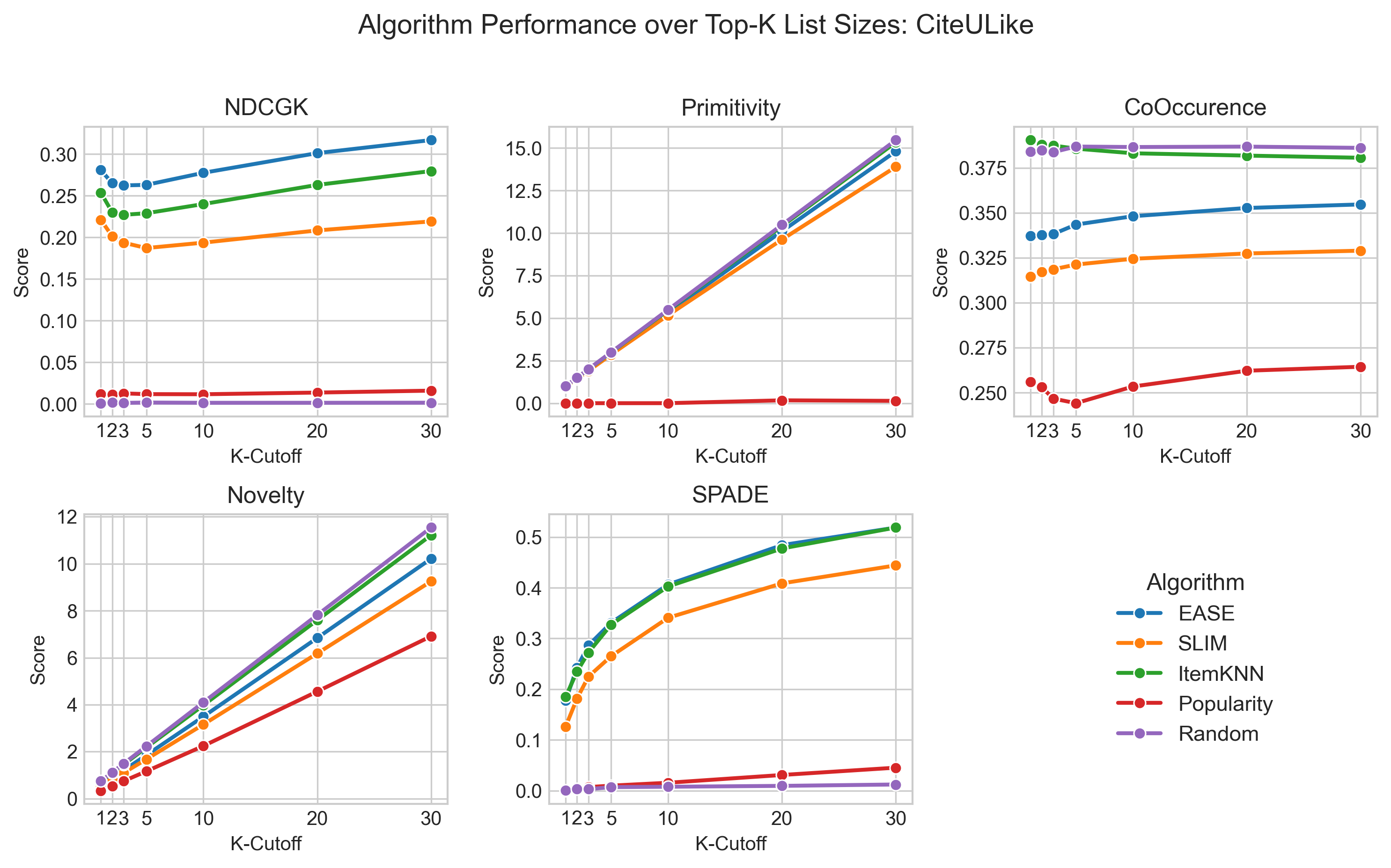}    
    \caption{Performance comparison of baseline recommendation algorithms (EASE, SLIM, ItemKNN, and Popularity) across varying recommendation list sizes ($K \in \{1, 2, 3, 5, 10, 20, 30\}$). The subplots depict the evaluation trajectories for accuracy (NDCG) and beyond-accuracy metrics (Primitivity, Co-Occurrence Surprise, Novelty, and SPADE) as the cutoff threshold increases.}
    \label{fig:my_stacked_images}
\end{figure}

\begin{table}[]
\centering
\begin{tabular}{llrrrrr}
\toprule
Dataset & Algorithm & NDCGK & Primitivity & CoOccurence & Novelty & SPADE \\
\midrule
Netflix & EASE & 0.315685 & 4.844650 & 0.103907 & 1.168139 & 0.127368 \\
Netflix & SLIM & 0.280544 & 4.544350 & 0.100814 & 1.110615 & 0.117351 \\
Netflix & ItemKNN & 0.190461 & 3.705450 & 0.085582 & 0.993255 & 0.061379 \\
Netflix & Popularity & 0.109781 & 0.000000 & 0.076752 & 0.783255 & 0.029467 \\
Netflix & Random & 0.001734 & 5.498050 & 0.214283 & inf & 0.004001 \\
\midrule
ML-1M & EASE & 0.334812 & 4.396189 & 0.095179 & 1.094719 & 0.101459 \\
ML-1M & SLIM & 0.299857 & 3.864540 & 0.089278 & 0.971579 & 0.085457 \\
ML-1M & ItemKNN & 0.258572 & 3.752361 & 0.085493 & 1.013478 & 0.068455 \\
ML-1M & Popularity & 0.149624 & 0.000000 & 0.068181 & 0.687975 & 0.030897 \\
ML-1M & Random & 0.007858 & 5.476885 & 0.179450 & 2.941511 & 0.014116 \\
\midrule
MIND & EASE & 0.528621 & 5.281100 & 0.160480 & 1.738990 & 0.234498 \\
MIND & SLIM & 0.468111 & 5.185100 & 0.156472 & 1.632805 & 0.210349 \\
MIND & ItemKNN & 0.360481 & 5.016600 & 0.151883 & 1.917122 & 0.194649 \\
MIND & Popularity & 0.063749 & 0.000000 & 0.116312 & 1.038346 & 0.036538 \\
MIND & Random & 0.000617 & 5.498997 & 0.281789 & inf & 0.003692 \\
\midrule
ML-20M & EASE & 0.313071 & 4.349293 & 0.116474 & 1.230299 & 0.131990 \\
ML-20M & SLIM & 0.277540 & 4.074596 & 0.114782 & 1.169247 & 0.115773 \\
ML-20M & ItemKNN & 0.238583 & 4.194949 & 0.108970 & 1.134467 & 0.101111 \\
ML-20M & Popularity & 0.127909 & 0.000000 & 0.087457 & 0.783185 & 0.031791 \\
ML-20M & Random & 0.001366 & 5.496224 & 0.221319 & inf & 0.002831 \\
\midrule
CiteULike & EASE & 0.277414 & 5.330776 & 0.348068 & 3.494691 & 0.406407 \\
CiteULike & SLIM & 0.193589 & 5.174007 & 0.324404 & 3.157389 & 0.340890 \\
CiteULike & ItemKNN & 0.239903 & 5.459838 & 0.383043 & 3.975786 & 0.402518 \\
CiteULike & Popularity & 0.011624 & 0.000903 & 0.253401 & 2.244291 & 0.015532 \\
CiteULike & Random & 0.001361 & 5.496300 & 0.386524 & 4.097694 & 0.007649 \\
\bottomrule
\end{tabular}
\caption{Cross-domain evaluation of baseline recommendation models (EASE, SLIM, ItemKNN, and Popularity) at a top-$K$ cutoff of 10. The table details ranking accuracy alongside beyond-accuracy metrics to illustrate the trade-off between relevance and serendipitous discovery across the different datasets.}
\label{tab:results}
\end{table}
Traditional metrics, such as Novelty, Primitivity, and Co-Occurrence, evaluate recommended items irrespective of their actual relevance to the user. 
Consequently, an algorithm can mathematically maximize traditional beyond-accuracy metric scores by simply recommending random, irrelevant items. 
The empirical evaluation clearly illustrates this vulnerability, e.g., on the MIND dataset, the Random algorithm yields a near-zero ranking accuracy (NDCG of 0.0006), yet achieves very high Primitivity (5.4989) and Co-Occurrence (0.281789) scores (\cref{tab:results}, \cref{fig:my_stacked_images}). 
In contrast, SPADE ensures relevance by calculating serendipity scores only for predicted items included in the test set.
By accurately identifying random recommendations as unconstrained noise, SPADE successfully penalizes this lack of relevance and assigns a proportionally low serendipity score of 0.0036 (\cref{tab:results}, \cref{fig:my_stacked_images}).

This strict relevance constraint also resolves a critical blind spot when evaluating high-performing models. 
While traditional metrics, e.g., show that EASE on ML-1M achieves both high ranking accuracy (NDCG of 0.3348) and high novelty (1.09, \cref{tab:results}), looking at these metrics in isolation cannot verify whether the novel recommendations are actually the ones the user found relevant.

Finally, SPADE successfully passes sanity checks by confirming that the non-personalized Random and Popularity baselines yield minimal serendipitous value. True serendipity inherently requires recommending relevant items that avoid both high similarity to the user's history and high global popularity. This expectation is perfectly mirrored by the Popularity and Random baselines, which consistently yield the lowest SPADE scores across all evaluated datasets (e.g., 0.0036 for Random on MIND, and 0.0294 for Popularity on Netflix).

\paragraph{Conclusion}
In conclusion, SPADE provides a robust, empirical framework for algorithm evaluation in recommender systems. 
By mathematically grounding serendipity in test-set relevance, SPADE prioritizes relevance and penalizes highly personalized or popular recommendations. 
Consequently, SPADE enables researchers to reliably distinguish genuinely serendipitous recommendations from those that simply exploit isolated beyond-accuracy metrics.

%% file: sample-ceur.bib
@inproceedings{10.1145/1864708.1864761,
author = {Ge, Mouzhi and Delgado-Battenfeld, Carla and Jannach, Dietmar},
title = {Beyond accuracy: evaluating recommender systems by coverage and serendipity},
year = {2010},
isbn = {9781605589060},
publisher = {Association for Computing Machinery},
address = {New York, NY, USA},
url = {https://doi.org/10.1145/1864708.1864761},
doi = {10.1145/1864708.1864761},
booktitle = {Proceedings of the Fourth ACM Conference on Recommender Systems},
pages = {257–260},
numpages = {4},
location = {Barcelona, Spain},
series = {RecSys '10}
}

@article{ziarani2021serendipity,
  title={Serendipity in recommender systems: a systematic literature review},
  author={Ziarani, Reza Jafari and Ravanmehr, Reza},
  journal={Journal of Computer Science and Technology},
  volume={36},
  number={2},
  pages={375--396},
  year={2021},
  publisher={Springer}
}

@inproceedings{10.1145/2792838.2800184,
author = {Maksai, Andrii and Garcin, Florent and Faltings, Boi},
title = {Predicting Online Performance of News Recommender Systems Through Richer Evaluation Metrics},
year = {2015},
isbn = {9781450336925},
publisher = {Association for Computing Machinery},
address = {New York, NY, USA},
url = {https://doi.org/10.1145/2792838.2800184},
doi = {10.1145/2792838.2800184},
booktitle = {Proceedings of the 9th ACM Conference on Recommender Systems},
pages = {179–186},
numpages = {8},
location = {Vienna, Austria},
series = {RecSys '15}
}

@article{10.1145/2926720,
author = {Kaminskas, Marius and Bridge, Derek},
title = {Diversity, Serendipity, Novelty, and Coverage: A Survey and Empirical Analysis of Beyond-Accuracy Objectives in Recommender Systems},
year = {2016},
issue_date = {March 2017},
publisher = {Association for Computing Machinery},
address = {New York, NY, USA},
volume = {7},
number = {1},
issn = {2160-6455},
url = {https://doi.org/10.1145/2926720},
doi = {10.1145/2926720},
journal = {ACM Trans. Interact. Intell. Syst.},
month = dec,
articleno = {2},
numpages = {42}
}

@inproceedings{10.1145/3167132.3167276,
author = {Kotkov, Denis and Konstan, Joseph A. and Zhao, Qian and Veijalainen, Jari},
title = {Investigating serendipity in recommender systems based on real user feedback},
year = {2018},
isbn = {9781450351911},
publisher = {Association for Computing Machinery},
address = {New York, NY, USA},
url = {https://doi.org/10.1145/3167132.3167276},
doi = {10.1145/3167132.3167276},
booktitle = {Proceedings of the 33rd Annual ACM Symposium on Applied Computing},
pages = {1341–1350},
numpages = {10},
location = {Pau, France},
series = {SAC '18}
}

@inproceedings{jenders2015serendipity,
  title={A serendipity model for news recommendation},
  author={Jenders, Maximilian and Lindhauer, Thorben and Kasneci, Gjergji and Krestel, Ralf and Naumann, Felix},
  booktitle={Joint German/Austrian Conference on Artificial Intelligence (K{\"u}nstliche Intelligenz)},
  pages={111--123},
  year={2015},
  organization={Springer}
}

@inproceedings{10.1145/3711896.3737199,
author = {Xi, Yunjia and Weng, Muyan and Chen, Wen and Yi, Chao and Chen, Dian and Guo, Gaoyang and Zhang, Mao and Wu, Jian and Jiang, Yuning and Liu, Qingwen and Yu, Yong and Zhang, Weinan},
title = {Bursting Filter Bubble: Enhancing Serendipity Recommendations with Aligned Large Language Models},
year = {2025},
isbn = {9798400714542},
publisher = {Association for Computing Machinery},
address = {New York, NY, USA},
url = {https://doi.org/10.1145/3711896.3737199},
doi = {10.1145/3711896.3737199},
booktitle = {Proceedings of the 31st ACM SIGKDD Conference on Knowledge Discovery and Data Mining V.2},
pages = {5059–5070},
numpages = {12},
location = {Toronto ON, Canada},
series = {KDD '25}
}

@inproceedings{10.1145/2566486.2568012,
author = {Nguyen, Tien T. and Hui, Pik-Mai and Harper, F. Maxwell and Terveen, Loren and Konstan, Joseph A.},
title = {Exploring the filter bubble: the effect of using recommender systems on content diversity},
year = {2014},
isbn = {9781450327442},
publisher = {Association for Computing Machinery},
address = {New York, NY, USA},
url = {https://doi.org/10.1145/2566486.2568012},
doi = {10.1145/2566486.2568012},
booktitle = {Proceedings of the 23rd International Conference on World Wide Web},
pages = {677–686},
numpages = {10},
location = {Seoul, Korea},
series = {WWW '14}
}

@inproceedings{10.1145/2124295.2124300,
author = {Zhang, Yuan Cao and S\'{e}aghdha, Diarmuid \'{O} and Quercia, Daniele and Jambor, Tamas},
title = {Auralist: introducing serendipity into music recommendation},
year = {2012},
isbn = {9781450307475},
publisher = {Association for Computing Machinery},
address = {New York, NY, USA},
url = {https://doi.org/10.1145/2124295.2124300},
doi = {10.1145/2124295.2124300},
booktitle = {Proceedings of the Fifth ACM International Conference on Web Search and Data Mining},
pages = {13–22},
numpages = {10},
location = {Seattle, Washington, USA},
series = {WSDM '12}
}

@incollection{shani2010evaluating,
  title={Evaluating recommendation systems},
  author={Shani, Guy and Gunawardana, Asela},
  booktitle={Recommender systems handbook},
  pages={257--297},
  year={2010},
  publisher={Springer}
}

@article{de2014comparison,
  title={Comparison of group recommendation algorithms},
  author={De Pessemier, Toon and Dooms, Simon and Martens, Luc},
  journal={Multimedia tools and applications},
  volume={72},
  number={3},
  pages={2497--2541},
  year={2014},
  publisher={Springer}
}

@article{wang2018serendipitous,
  title={Serendipitous recommendation in e-commerce using innovator-based collaborative filtering},
  author={Wang, Chang-Dong and Deng, Zhi-Hong and Lai, Jian-Huang and Yu, Philip S},
  journal={IEEE transactions on cybernetics},
  volume={49},
  number={7},
  pages={2678--2692},
  year={2018},
  publisher={IEEE}
}

@inproceedings{murakami2007metrics,
  title={Metrics for evaluating the serendipity of recommendation lists},
  author={Murakami, Tomoko and Mori, Koichiro and Orihara, Ryohei},
  booktitle={Annual conference of the Japanese society for artificial intelligence},
  pages={40--46},
  year={2007},
  organization={Springer}
}

@inproceedings{kaminskas2014measuring,
  title={Measuring surprise in recommender systems},
  author={Kaminskas, Marius and Bridge, Derek},
  booktitle={Proceedings of the workshop on recommender systems evaluation: dimensions and design (Workshop programme of the 8th ACM conference on recommender systems)},
  year={2014},
  organization={Citeseer}
}

@article{kaminskas2016diversity,
  title={Diversity, serendipity, novelty, and coverage: a survey and empirical analysis of beyond-accuracy objectives in recommender systems},
  author={Kaminskas, Marius and Bridge, Derek},
  journal={ACM Transactions on Interactive Intelligent Systems (TiiS)},
  volume={7},
  number={1},
  pages={1--42},
  year={2016},
  publisher={ACM New York, NY, USA}
}

@inproceedings{kawamae2015real,
  title={Real time recommendations from connoisseurs},
  author={Kawamae, Noriaki},
  booktitle={Proceedings of the 21th ACM SIGKDD international conference on knowledge discovery and data mining},
  pages={537--546},
  year={2015}
}

@incollection{robbins1992empirical,
  title={An empirical Bayes approach to statistics},
  author={Robbins, Herbert E},
  booktitle={Breakthroughs in Statistics: Foundations and basic theory},
  pages={388--394},
  year={1992},
  publisher={Springer}
}

@inproceedings{Bouma2009NormalizedM,
  title={Normalized (pointwise) mutual information in collocation extraction},
  author={Gerlof J. Bouma},
  year={2009},
  url={https://api.semanticscholar.org/CorpusID:2762657}
}

@article{jelassi2015towards,
  title={Towards more targeted recommendations in folksonomies},
  author={Jelassi, Mohamed Nader and Ben Yahia, Sadok and Mephu Nguifo, Engelbert},
  journal={Social Network Analysis and Mining},
  volume={5},
  number={1},
  pages={68},
  year={2015},
  publisher={Springer}
}

@inproceedings{chen2021values,
  title={Values of user exploration in recommender systems},
  author={Chen, Minmin and Wang, Yuyan and Xu, Can and Le, Ya and Sharma, Mohit and Richardson, Lee and Wu, Su-Lin and Chi, Ed},
  booktitle={Proceedings of the 15th acm Conference on recommender systems},
  pages={85--95},
  year={2021}
}

@inproceedings{sa2022diversity,
  title={Diversity vs relevance: A practical multi-objective study in luxury fashion recommendations},
  author={S{\'a}, Jo{\~a}o and Queiroz Marinho, Vanessa and Magalh{\~a}es, Ana Rita and Lacerda, Tiago and Goncalves, Diogo},
  booktitle={Proceedings of the 45th International ACM SIGIR Conference on research and development in information retrieval},
  pages={2405--2409},
  year={2022}
}

@inproceedings{maksai2015predicting,
  title={Predicting online performance of news recommender systems through richer evaluation metrics},
  author={Maksai, Andrii and Garcin, Florent and Faltings, Boi},
  booktitle={Proceedings of the 9th ACM Conference on Recommender Systems},
  pages={179--186},
  year={2015}
}

@article{kotkov2017serendipity,
  title={A serendipity-oriented greedy algorithm for recommendations},
  author={Kotkov, Denis and Veijalainen, Jari and Wang, Shuaiqiang and Majchrzak, Tim A and Traverso, Paolo and Krempels, Karl-Heinz and Monfort, Val{\'e}rie},
  year={2017},
  publisher={SCITEPRESS Science And Technology Publications}
}

@inproceedings{babu2015evaluation,
  title={Evaluation of similarity based recommender using Fuzzy Inference System},
  author={Babu, LD Dhinesh and Raj, Ebin Deni and others},
  booktitle={2015 International Conference on Computing and Network Communications (CoCoNet)},
  pages={576--581},
  year={2015},
  organization={IEEE}
}

@inproceedings{nishioka2019towards,
  title={Towards serendipitous research paper recommender using tweets and diversification},
  author={Nishioka, Chifumi and Hauk, J{\"o}rn and Scherp, Ansgar},
  booktitle={International Conference on Theory and Practice of Digital Libraries},
  pages={339--343},
  year={2019},
  organization={Springer}
}

@article{nishioka2020influence,
  title={Influence of tweets and diversification on serendipitous research paper recommender systems},
  author={Nishioka, Chifumi and Hauke, J{\"o}rn and Scherp, Ansgar},
  journal={PeerJ Computer Science},
  volume={6},
  pages={e273},
  year={2020},
  publisher={PeerJ Inc.}
}

@article{ren2015survey,
  title={A survey of recommendation techniques based on offline data processing},
  author={Ren, Yongli and Li, Gang and Zhou, Wanlei},
  journal={Concurrency and Computation: Practice and Experience},
  volume={27},
  number={15},
  pages={3915--3942},
  year={2015},
  publisher={Wiley Online Library}
}

@article{adamopoulos2014unexpectedness,
  title={On unexpectedness in recommender systems: Or how to better expect the unexpected},
  author={Adamopoulos, Panagiotis and Tuzhilin, Alexander},
  journal={ACM Transactions on Intelligent Systems and Technology (TIST)},
  volume={5},
  number={4},
  pages={1--32},
  year={2014},
  publisher={ACM New York, NY, USA}
}

@article{ziarani2021deep,
  title={Deep neural network approach for a serendipity-oriented recommendation system},
  author={Ziarani, Reza Jafari and Ravanmehr, Reza},
  journal={Expert Systems with Applications},
  volume={185},
  pages={115660},
  year={2021},
  publisher={Elsevier}
}

@inproceedings{kawamae2010serendipitous,
  title={Serendipitous recommendations via innovators},
  author={Kawamae, Noriaki},
  booktitle={Proceedings of the 33rd international ACM SIGIR conference on Research and development in information retrieval},
  pages={218--225},
  year={2010}
}

@article{kaya2022novel,
  title={A novel top-n recommendation method for multi-criteria collaborative filtering},
  author={Kaya, Tugba and Kaleli, Cihan},
  journal={Expert Systems with Applications},
  volume={198},
  pages={116695},
  year={2022},
  publisher={Elsevier}
}

@inproceedings{lin2019pareto,
  title={A pareto-efficient algorithm for multiple objective optimization in e-commerce recommendation},
  author={Lin, Xiao and Chen, Hongjie and Pei, Changhua and Sun, Fei and Xiao, Xuanji and Sun, Hanxiao and Zhang, Yongfeng and Ou, Wenwu and Jiang, Peng},
  booktitle={Proceedings of the 13th ACM Conference on recommender systems},
  pages={20--28},
  year={2019}
}

@article{ribeiro2014multiobjective,
  title={Multiobjective pareto-efficient approaches for recommender systems},
  author={Ribeiro, Marco Tulio and Ziviani, Nivio and Moura, Edleno Silva De and Hata, Itamar and Lacerda, Anisio and Veloso, Adriano},
  journal={ACM Transactions on Intelligent Systems and Technology (TIST)},
  volume={5},
  number={4},
  pages={1--20},
  year={2014},
  publisher={ACM New York, NY, USA}
}

@article{li2024deep,
  title={Deep pareto reinforcement learning for multi-objective recommender systems},
  author={Li, Pan and Tuzhilin, Alexander},
  journal={arXiv preprint arXiv:2407.03580},
  year={2024}
}

@inproceedings{xiao2017fairness,
  title={Fairness-aware group recommendation with pareto-efficiency},
  author={Xiao, Lin and Min, Zhang and Yongfeng, Zhang and Zhaoquan, Gu and Yiqun, Liu and Shaoping, Ma},
  booktitle={Proceedings of the eleventh ACM conference on recommender systems},
  pages={107--115},
  year={2017}
}

@inproceedings{peng2024reconciling,
  title={Reconciling the accuracy-diversity trade-off in recommendations},
  author={Peng, Kenny and Raghavan, Manish and Pierson, Emma and Kleinberg, Jon and Garg, Nikhil},
  booktitle={Proceedings of the ACM Web Conference 2024},
  pages={1318--1329},
  year={2024}
}

@article{10.1145/2827872,
author = {Harper, F. Maxwell and Konstan, Joseph A.},
title = {The MovieLens Datasets: History and Context},
year = {2015},
issue_date = {January 2016},
publisher = {Association for Computing Machinery},
address = {New York, NY, USA},
volume = {5},
number = {4},
issn = {2160-6455},
url = {https://doi.org/10.1145/2827872},
doi = {10.1145/2827872},
journal = {ACM Trans. Interact. Intell. Syst.},
month = {dec},
articleno = {19},
numpages = {19}
}

@inproceedings{DBLP:conf/ijcai/WangCL13,
  author    = {Hao Wang and
               Binyi Chen and
               Wu-Jun Li},
  title     = {Collaborative Topic Regression with Social Regularization
               for Tag Recommendation},
  booktitle = {IJCAI},
  year      = {2013}
}

@inproceedings{wu2020mind,
  title={Mind: A large-scale dataset for news recommendation},
  author={Wu, Fangzhao and Qiao, Ying and Chen, Jiun-Hung and Wu, Chuhan and Qi, Tao and Lian, Jianxun and Liu, Danyang and Xie, Xing and Gao, Jianfeng and Wu, Winnie and others},
  booktitle={Proceedings of the 58th annual meeting of the association for computational linguistics},
  pages={3597--3606},
  year={2020}
}

@inproceedings{michiels2022recpack,
  title={Recpack: An (other) experimentation toolkit for top-n recommendation using implicit feedback data},
  author={Michiels, Lien and Verachtert, Robin and Goethals, Bart},
  booktitle={Proceedings of the 16th ACM Conference on Recommender Systems},
  pages={648--651},
  year={2022}
}

@inproceedings{barkan2021anchor,
  title={Anchor-based collaborative filtering},
  author={Barkan, Oren and Hirsch, Roy and Katz, Ori and Caciularu, Avi and Koenigstein, Noam},
  booktitle={Proceedings of the 30th ACM International Conference on Information \& Knowledge Management},
  pages={2877--2881},
  year={2021}
}

@inproceedings{melchiorre2022protomf,
  title={Protomf: Prototype-based matrix factorization for effective and explainable recommendations},
  author={Melchiorre, Alessandro B and Rekabsaz, Navid and Ganh{\"o}r, Christian and Schedl, Markus},
  booktitle={Proceedings of the 16th ACM Conference on Recommender Systems},
  pages={246--256},
  year={2022}
}

@article{lin2024recommendation,
  title={How do recommendation models amplify popularity bias? An analysis from the spectral perspective},
  author={Lin, Siyi and Gao, Chongming and Chen, Jiawei and Zhou, Sheng and Hu, Binbin and Feng, Yan and Chen, Chun and Wang, Can},
  journal={arXiv preprint arXiv:2404.12008},
  year={2024}
}
